\documentclass[preprint2]{aastex}

\slugcomment{{\it Astronomical Journal}, in press, September 2001}

\begin{document}

\title{Seeing Galaxies though Thick and Thin.
\\ IV. The Superimposed Spiral Galaxies of NGC 3314\altaffilmark{1}}

\author{William C. Keel\altaffilmark{2,3} and Raymond E. White III}
\affil{Department of Physics and Astronomy, Box 870324, University of Alabama,
Tuscaloosa, AL 35487}

\altaffiltext{1}{Based on observations with the NASA/ESA 
{\it Hubble Space Telescope} obtained at the Space Telescope Science Institute, 
which is operated by the Association of Universities for Research in Astronomy, 
Inc., under NASA contract No. NAS5-26555.}
\altaffiltext{2}{Visiting astronomer, NASA Infrared Telescope Facility,
Mauna Kea, Hawaii}
\altaffiltext{3}{Visiting Astronomer, WIYN Observatory, which is owned and 
operated by 
the WIYN Consortium, Inc., which consists of the University of Wisconsin, 
Indiana University, Yale University, and the National Optical Astronomy 
Observatory (NOAO). NOAO is operated for the National Science Foundation 
by the Association of Universities for Research in Astronomy (AURA), Inc.}

\begin{abstract}
The superimposed pair of spiral galaxies
comprising NGC 3314 offers a unique opportunity to trace
the dust properties in a spiral galaxy.
We analyze multicolor HST imaging, supported by ground-based near-IR imaging 
and fiber-array spectroscopy to measure dust extinction in 
the foreground Sc galaxy NGC 3314A, which is backlit by 
the Sb system NGC 3314B.  The superposition allows us to measure
extinctions over a wide range of galactocentric radii
in the foreground galaxy, from 0.4-4.5 kpc.

In the outer half of the disk, the extinction is strongly localized in discrete 
dust lanes, including some patches whose galactic setting is clear
only because of associated H$\alpha$ emission at the foreground
velocity. These dust features show an extinction curve with a slope close to 
the Galactic mean ($R = 3.5 \pm 0.3$) over a range in
galactocentric radius from 1.6 to 3.8 kpc, with no radial trend.
Using the $I-K$ color of the background nucleus, we derive an extinction 
$A_I = 3.3$ through the disk at a projected distance 400 pc 
from the nucleus of NGC 3314A.

The extinction in even the inner disk of NGC 3314A is quite patchy,
since background H$\alpha$ emission is detected from all parts of
the system. Local anticorrelations between foreground and
background line emission demonstrate that the dust is concentrated to
star-forming regions, as has been found for the blue light in several systems.

Colors of dust lanes in NGC 3314A which are projected only partially
against the background disk indicate that the dust scale height
in the foreground disk is substantially smaller than that of the
stars. The color-intensity behavior of the net light in these
regions tracks the predictions of a thin-layer model closely.

\end{abstract}


\section{Introduction}

Properly accounting for the effects of dust upon light from galaxies 
is crucial to our understanding of galaxy evolution,
of the interplay between stars and interstellar medium within galaxies,
of the most distant galaxies (whose UV radiation is
mostly self-absorbed), and of the 
cumulative absorption from foreground galaxies which may affect
the observability of high-redshift QSOs.
Galaxy superpositions have recently been exploited to directly measure 
the dust content and distribution within galaxies, free of many of the
assumptions about internal structure required by techniques that
use only a galaxy's own light. We have pursued this approach
in a series of papers, successively refining its scope for
mapping dust within galaxies. White \& Keel (1992) used
ground-based imaging of AM1316-241 to map the extinction in and around a 
prominent spiral arm, showing that interarm extinction can be small.
Using a larger sample of 12 suitable backlit spirals, White, Keel,
\& Conselice (2000; WKC) found this to be representative behavior;
arms and resonance rings can have substantial opacities 
($\tau_B > 1$) at any radius,
while interarm extinction shows a roughly exponential radial
distribution and reaches $A_B = 0.5$ only within about $0.4 R_{25}$.
These results also showed extinction curves greyer than the Galactic mean,
indicative of unresolved dust structure. This was borne out
by HST imaging of AM0500-620 and AM1316-241 (Keel \& White 2001), for
which the extinction behavior is much like the Milky Way's when 
measured at resolutions of 30 pc. Similar behavior was found for
dust lanes in the outer disk of NGC 2207 projected in front of its ocular
companion IC 2163 (Elmegreen et al. 2001).
Combining dust masses estimated from optical
absorption in backlit spirals with ISO and JCMT infrared data to estimate
thermal-emission masses, Domingue et al. (1999) found that these estimates
are in good agreement, limiting the role of very cold dust or
small dense clumps. Domingue et al. (2000) added to the sample of
workable galaxies for these studies by using spectroscopy to split the
light from overlapping pairs of spirals, based on their redshift
differences, finding concordance with results from spiral-elliptical pairs.

While it violates some of our precepts for a full analysis using the partial
overlap technique, the unique galaxy superposition NGC 3314 allows
us to probe extinction closer to the nucleus of a spiral
galaxy than in any other nearby galaxy pair. It also allows us to abandon 
the symmetry contraints that make 
grand-design spirals the most amenable to our earlier partial
overlap analyses, since NGC 3314A has its spiral pattern outlined
as much by the dust pattern as by bright stellar associations (as
noted by Sandage 1961 for NGC 253). NGC 3314, located in the direction
of the Hydra I cluster (Abell 1060), consists of an Sc galaxy (NGC3314A)
in front of an earlier-type Sb system (NGC 3314B), with their nuclei separated
by only $2.36^{\prime\prime}$. The line-of-sight separation is poorly known, 
since
both galaxies have redshifts within the range of the velocity dispersion
of Abell 1060. We derive heliocentric
central values of $cz=$2834 and 4665 km s$^{-1}$
for NGC 3314A and NGC 3314B respectively, in good agreement with the
values found from slit spectroscopy by Schweizer \& Thonnard (1985). 
There are additional galaxies, including the giant spiral
NGC 3312, at redshifts very close to NGC 3314A, and some evidence for
H I tails indicates that these galaxies form a distinct, weakly interacting
group in the foreground of Abell 1060 (McMahon et al. 1992). If so, this group's
redshift distance would place it at about 35 Mpc, while the mean
redshift of 3400 km s$^{-1}$ for Abell 1060 puts its core at 42
Mpc (we take H$_0 = 80$ km s$^{-1}$ Mpc$^{-1}$ throughout). The core
radius of Abell 1060 is much smaller than this difference of
7 Mpc, so that essentially any separation ranging from small values to
7 Mpc is possible. The lack of substantial distortion in the outer
isophotes of NGC 3314B suggests that the two systems are separated
by at least the sum of their radii to detectable surface-brightness
levels. This matters mostly in evaluating the possible contribution
of scattered background light to what we measure, as outlined in the
appendix of WKC. The velocity field in H$\alpha$ will allow us to
place limits on the role of scattering in the light reaching
us from NGC 3314B.

The dust in NGC 3314A has been examined previously, using ground-based imaging,
by Keel (1983) and WKC, and via spectroscopy
to look for excess Balmer decrement by James \& Puxley (1993). The improvement
in structure discrimination, and more accurate recovery of the extinction
curve, shown by comparing ground-based and HST observations of 
spirals backlit by ellipticals (Keel \& White 2001) suggested that
similarly spectacular gains could be realized for the entire dust pattern
in NGC 3314A. We describe here the results of multicolor HST WFPC2
imaging of NGC 3314, supplemented by ground-based $K$ imaging and
spectroscopic mapping, which not only show the reddening
curve of material in the Sc foreground galaxy, but let us derive the
extinction within half a kiloparsec of its nucleus. Radial locations
of features have been derived with a geometrical model based on
the new H$\alpha$ velocity fields, which gives parameters for 
NGC 3314A intermediate between the results from optical rotation
curves by Schweizer \& Thonnard (1985) and the H I velocity field
by McMahon et al. (1992). We take an inclination of $i = 41 \pm 1 ^\circ$
to the line of sight, and a major axis position angle of $20^\circ$
(as set out in section 2.3). For conversion to linear radius within
NGC 3314A, we adopt a distance of 36 Mpc, or a scale of 172 pc/arcsecond.

\section{Observations}

\subsection{HST imaging}

We obtained F450W ($B$) and F814W ($I$) images of NGC 3314 under GO program
6438, using WFPC2
on 5 April 1999, with total exposure times 2600 and 2400 seconds respectively,
split into halves for cosmic-ray rejection. These filters were
selected, as in Keel \& White (2001), to give a long
wavelength baseline with high quantum efficiency and without including
strong emission lines. The orientation 
(U3 position angle 84.9$^\circ$) was
specified so as to place a bright star, superimposed on the
northwest end of NGC 3314B, on another CCD so that charge bleeding or
scattered starlight
would not compromise the galaxies' image. We deal here with data from the
WF3 CCD, including the whole region of dust backlighting. The B image was
supplemented with data from the Hubble Heritage observations obtained
on 10 March 2000, adding 1400 seconds of exposure in a cosmic-ray split
pair. The orientation was set to be identical for this second
observation, and indeed no rotation was needed to combine the images.
An offset of 0.2 pixel in each coordinate was handled by interpolation
of the shorter Heritage F450W exposure, introducing a small smoothing which
has no consequence to our analysis (since the PSF is significantly larger in 
F814W). Fig. 1 shows a color composite from
both data sets, including as well F555W and F675W images taken for the Heritage
project (mapped as green together).

\subsection{Near-Infrared Imaging}

We use a $K$-band image obtained with the NSFCam imager at the
3-m NASA Infrared Telescope Facility on Mauna Kea. The camera was
set to deliver $0.30^{\prime\prime}$ pixels for a $77^{\prime\prime}$ field, 
with nearby blank sky used as
a flat-field reference and the final mosaic constructed from 
$3 \times 3$ dither patterns with $20^{\prime\prime}$ spacing between pointings.
The total exposure on NGC 3314 was 18 minutes. Photometric calibration
was based on UKIRT faint standard stars FS21 and FS27 observed before
and after NGC 3314. Star images show a resolution of $1.0^{\prime\prime}$ FWHM
for the final image stack.
The final processed image is shown in Fig. 2. With the much better
penetration of $2.2 \mu$ photons, the background nucleus and both
spirals patterns become evident.

\subsection{Fiber-array Spectroscopy}

The spectral region including H$\alpha$, [N II] $\lambda \lambda 6548,6584$, 
and [S II] $\lambda \lambda 6717,6731$ was mapped using the DensePak
fiber array (Barden, Sawyer, \& Honeycutt 1998) at 
the 3.5m WIYN telescope during two nights in December 2000.
The array has fibers with $3^{\prime\prime}$ aperture diameter, arranged in a
$7 \times 13$ configuration, nominally in hexagonal packing 
on a $4^{\prime\prime}$ triangular grid. Sky subtraction uses four 
outlying fibers, located about $60^{\prime\prime}$ from the array 
center in diagonal directions with respect to the rectangular array outline.
The major axis of the array was set north-south. The full
spectral range covered was 6010-7410 \AA\  at 0.69 \AA\  per pixel,
sampling an effective resolution of 2.0 pixels or 1.4 \AA\  FWHM.
Seven pointings were obtained. One set of four 30-minute exposures
(with the last truncated at 20 because of twilight) was taken
in a trapezoidal pattern, parallel to the array packing to fill the
fiber gaps. In addition, a set of 45-minute exposures were
obtained centered on the nucleus and at positions offset $20^{\prime\prime}$ 
in each
coordinate to the northwest and southeast, roughly along the major
axis of NGC 3314B, to complete the kinematic coverage of the system.
Reduction included extraction of the fiber spectra to one-dimensional
form, with S/N ratio optimized by variance weighting perpendicular to
the dispersion, and cosmic-ray rejection based on the CCD noise parameters.
The spectra were rebinned to a common linear wavelength scale to allow
sky subtraction. Line wavelengths were measured for each fiber position
via the IRAF {\tt splot} task, by Gaussian fitting.
 
Two-dimensional maps of emission-line wavelength and intensity were
constructed from the fiber measurements, by averaging data from all overlapping
positions on a subpixel numerical grid averaged to a pixel spacing of
$1^{\prime\prime}$. 
Velocity fields for the two galaxies separately are shown in
Fig. 3 for NGC 3314A and Fig. 4 for NGC 3314B.

Each set of fiber observations began with a pointing visually centered on 
the nucleus of NGC 3314A in a selected fiber aperture, using a TV view with the
fibers back-illuminated. This introduces about $1^{\prime\prime}$ uncertainty 
in the
positioning of the two nights' data against each other and with respect
to the images. The various pointings were registered by requiring
minimal discontinuities in the velocity field near the center, where
the highest-amplitude structure is found. Registration against the HST images
assumes that the peak brightness of the nucleus of NGC 3314A occurs
when a fiber is geometrically centered on it.

We have derived kinematic estimates for the inclination and orientation
of each disk from these data. Using the gridded velocity maps,
we applied the approach of Warner, Wright, \& Baldwin (1973)
as implemented in
code provided by R. Buta, in which the geometric parameters (disk inclination,
major-axis orientation, central velocity, and center location) are varied 
iteratively to 
minimize the scatter in the implied (deprojected) rotation curve.
For NGC 3314A, we derive an inclination of $41 \pm 1^\circ$
(where $90^\circ$ is edge-on), major axis position angle
$20 \pm 1 ^\circ$, and central velocity $2809 \pm 2$ km s$^{-1}$
as observed (2834 corrected to heliocentric). The rotation curves are shown in 
Fig. 5. For the background
system NGC 3314B, we derive an inclination of $63 \pm 1^\circ$,
position angle of $142 \pm 1^\circ$, and observed central velocity
$cz = 4640 \pm 2$ km s$^{-1}$ (4665 heliocentric). The 
best-fitting kinematic center is significantly 
offset from the near-IR nucleus, lying $1^{\prime\prime}$ west 
and $3.5^{\prime\prime}$ south of
the nucleus of NGC 3314A. Part of this offset may result from a bias in
the distribution of fitted points, which fill the area to the south more
fully and may be affected in particular by the strong extinction right at
the foreground nucleus. Adopting these orientations, we extract
the individual rotation curves as shown in Fig. 5. They are close
matches for the slices shown by Schweizer \& Thonnard 1985, with the
proviso that the central regions in our data will be affected by the limited 
spatial resolution afforded by $3^{\prime\prime}$ fibers and by patchy
sampling in NGC3314B.

The small scatter for points in the derived rotation curve of NGC 3314B,
away from the nucleus so that patchy sampling or exact centering are
not problems, indicates that scattering is not a problem in measuring
extinction. Scattered light, presumably strongly variable in intensity
as the dust column density changes, would introduce both point-to-point
scatter and an overall broadening of the H$\alpha$ profile, as light
from various parts of NGC 3314B would contribute to the observed
spectrum at each point.

\section{Distribution of Extinction in NGC 3314A}

We apply four techniques to measure the extinction through NGC 3314A
in various regimes. While independent of many of the model assumptions
used to measure extinction within a single galaxy, each of these does
have certain built-in geometric assumptions, which we summarize as follows.

Extinction in the prominent {\bf dust lanes} can be measured with respect
to their immediate surroundings (section 3.1). This uses
interpolation of the background light across the dust lane and a
correction for foreground light, derived from appropriate parts of the relevant
spiral pattern as seen against empty space. The only major
assumptions here are that background structures are not fortuitously correlated 
with foreground dust features, and that foreground light is not
so distributed as to mask true extinction (and in a way not seen
in ``clean'' foreground arms).

The {\bf interlane extinction} can be measured in the outer disk
of NGC 3314A, where we see structure in the background disk and thus can be
confident that the region is of low extinction. For these, we use the
$B-I$ color after correcting for the estimated foreground light, and
apply the same slope of the reddening law as found for the discrete
dust lanes. The assumptions going into these measurements are that
the interarm dust is not very different from that in the dust lanes
(so we can convert $B-I$ excess into $A_B$), and that the foreground
arms seen without backlight are fair samples of their behavior at
intermediate, backlit points.

We use H$\alpha$ emission from the background system to estimate the
effective transmission in the {\bf inner disk}. This can be done only
roughly, using the typical correlation between scale lengths in H$\alpha$
and the far-red continuum, as noted by Ryder \& Dopita (1994). This gives net
transmissions averaged over the $3^{\prime\prime}$ scale of our spectroscopic 
mapping,
and the errors will be larger than for the other methods because
H$\alpha$ will exhibit fine structure that we have no independent
way to recover from these data. Given the rather surprising prevalence of 
detected background H$\alpha$ emission, this technique could be substantially
improved by mapping in one of the near-IR recombination features
such as Pa$\alpha$ or Br$\gamma$.

The reddening toward the {\bf nucleus} of NGC 3314B can be estimated 
from its color excess in $I-K$, based on other spirals of similar
Hubble type. This assumes, of course, that the color behavior is reasonably
well-behaved for various galaxies, which is true for colors this
red, and that we can accurately measure both $I$ and $K$ fluxes in
the presence of large and very patchy extinction.

One important goal of these observations is to combine these techniques
to give a nearly complete radial profile of the extinction, both
in and between the dusty arms.
Our direct extinction measurements will in general be greater than the 
face-on extinction through the disk (by a factor of 1.33 for our adopted 
geometry, the secant of the inclination to the line of sight).
The detailed distribution of the dust affects this factor somewhat;
a uniform, vertically stratified layer will obey it exactly, while
a collection of isolated clouds sparsely filling the disk volume
will follow this behavior in covering factor rather than extinction.

\subsection{Interlane Extinction}

Lacking a direct comparison of superimposed and clear views of the
two galaxies, we must use more indirect means to estimate extinction
in the interlane regions (where we use ``interlane" to indicate 
that we are looking between the
dust lanes, not necessarily between optically bright spiral features). 
Such regions can be identified in the outer part
of NGC 3314A's disk because background spiral structure can be seen
through them, except for a small area northwest of its nucleus where 
the directions of both patterns coincide. Since the dust in the dark
lanes follows the Milky Way reddening law closely (section 4), we will
not be far off in taking this form to derive the (much smaller)
interarm extinction from color excess. We look for a color
gradient in interlane regions, which are identified from visibility of 
background
structure and lack of dust lanes following the foreground spiral
pattern. Assuming the background disk to have constant color
with radius should give an upper limit on foreground reddening (hence
extinction) since very few Sb galaxies become bluer at smaller radii
within the disk. Our best estimate is derived from the slopes of 
typical color gradients in Sb systems, over the $B-I$ baseline
which corresponds closely to the WFPC2 filters we used. We incorporate 
gradient data
from 14 Sb galaxies (excluding strongly barred systems) from 
Cunow (1998) and Aguerri et al. (2000). We measure the color gradients
across the radial range relevant for NGC 3314A, 
in the range 0.3--0.6$R_{25}$ (where $R_{25}$ is the radius to the
$B=25$ mag arcsec$^{-2}$ isophote used for the usual $D_{25}$ diameter
measurement). With $R_{25}$ for each system as the
unit of length, the gradients range from $\Delta (B-I) = -0.05$
for NGC 6970 to +1.53 for NGC 7753 (where a positive gradient
means the central regions are redder). Assuming flat color behavior
will then give an upper limit to the color excess in NGC 3314A, while
we can rule out the stronger gradients from the observed colors themselves.

The regions selected for interlane measurements are shown in Fig 6. We
attempted to find sets of regions that follow obvious color structure
in the disk of NGC 3314B, avoiding its dust lanes, so the color 
comparison among interlane regions would be more accurate. Foreground
intensity corrections use measurements of the spiral features in NGC 3314A
seen off the background disk, as close as possible to the position of the
interlane regions we measure and interpolated along the arm pattern
to account for their changing brightness with distance from the foreground
nucleus. The interlane regions specifically excluded bright clusters
in NGC 3314B, to avoid the possibility of bias toward unusually bright
and blue regions, which means that we always require the regions to show
background structure over areas more than $1^{\prime\prime}$ in size. We 
track errors
based on the fluctuations seen in the off-disk foreground regions
and the amount of interpolation needed between these regions to
estimate the foreground light at each interlane position.
We measured
mean values of $B/I$ count rate ratio in tailored polygonal regions for
each, converted into color excess $E(B-I)$ by comparison with the ratio
from the outer unobscured disk on NGC 3314B. These results are listed,
with computed galactocentric distances for each region
R(fg) and R(bg) for foreground and
background systems, in Table 1.
Here and in Table 2, the regions are listed with coordinate offsets
from the nucleus of NGC 3314A, which has a position in the GSC
reference frame (2000) 10$^{\rm h}$ 37$^{\rm m}$ 12$^{\rm s}$.900 
-27$^\circ$ 41$^\prime$ 01$^{\prime \prime}$.33.
Since the orientations of the two disks are different, we could in 
principle solve simultaneously for a background color gradient and
foreground reddening gradient. However, given the distribution of the
interlane regions we could identify, projected radii in both
disks are strongly coupled for the whole sample. They range from
1.7--4.1 kpc in the foreground disk and 2.5--5.5 kpc in the background
system. However, since we expect a genuine color gradient in NGC 3314B
to be monotonic, a limit on such a gradient can be set from the
majority of $(B-I)$ values which form a systematic trend after 
rejecting the three highest values (which do not occur among the
three points at the smallest radii, as shown in Fig. 7). This suggests 
a gradient of
$\Delta (B-I) = 0.03$ mag $^{-1}$, reddening inwards, corresponding
to about $\Delta(B-I) = 0.3$ over $R_{25}$. This places NGC 3314B
among the spirals of similar Hubble type with the flattest disk color
behavior.

We can use this estimate (formally an upper limit) of the background color 
gradient to interpret the color data in the interarm regions, which
is important not only for the interlane extinction itself but as a zero
level for the differential extinction in the prominent dust lanes.
A rough fit to the run of $(B-I)$ versus projected radius in NGC 3314A
(Fig. 7)
gives a slope of 0.12 mag kpc$^{-1}$, flattening to essentially zero for radii 
greater than 4 kpc. Since this color behavior is strongly affected by
how we interpret the outer interlane region at 3.8 kpc, we can
estimate a mean extinction from the H I column density reported by
McMahon et al. (1992), which is about $1.1 \times 10^{20}$ cm$^{-2}$
at these location at a resolution of $17 \times 30$" delivered by
their hybrid VLA configuration. A typical Galactic ratio of extinction
to H I column density would predict a value $A_V \approx 0.05$
from this average, consistent with out extrapolation.
There is substantial scatter in the radial color behavior, much of which 
is surely driven by the errors in assessing the correction for foreground light
in each region as well as genuine scatter among various
parts of the disk. Since the dust in the strong lanes has a reddening curve 
close to the Milky Way mean, we will not err much in assigning a similar
slope $R=3.1$ to the milder interarm extinction. For extinctions small
enough to keep the effective filter wavelengths constant, this means
$A_B = 2.12 E_{B-I}$, giving an interarm extinction behavior over
the range $R=1.8-4$ kpc
rising by $A_B= 0.25$ mag kpc$^{-1}$ going inward, starting from
zero outside of $R=4$ kpc. This is in reasonable accord with the
ground-based results of WKC, and for this first time lets us
evaluate interlane dust in a more flocculent spiral system
than could be analyzed before.
  
\subsection{Outer Dust Lanes}

For dust lanes $\approx 2$ kpc or more from the center of NGC 3314A, the 
effects of interlane
extinction and foreground structure are small enough that we can
measure extinction in the dust lanes differentially with respect
to their surroundings. This was done for nine discrete dust features
surrounded by interlane regions, as identified in Fig. 6 and
listed in Table 2.
We construct local models of the background
light by interpolation along the pitch angle of the background spiral
structure, and divide the data by these models to get maps of
transmitted intensity at each wavelength, B and I. This process is
illustrated in Fig. 8 for a typical dust lane. We do this with
two extreme values for correction to account for light from the
foreground disk in the dust lanes - zero as a lower limit, and assuming 
that the diffuse bright regions of the same arms when seen off the background
disk are representative. The latter is an upper limit, since the dust lanes
are also seen against NGC 3314A's own light when not backlit, and
we cannot do a more targeted correction (incorporating lane/interlane
distinctions in the pure background ares, for example) without
to some extent presuming the answer, since there is not a clear
way to make the same kind of distinctions in different illuminated
dust regions. This
largest correction in fact leads to
unphysical results (negative intensities over large areas) for
dust lanes found 1.7--1.8 kpc from the nucleus. Histograms of the
derived transmissions $T = e^{- \tau}$ in B for these dust 
lanes are shown in Fig. 9,
superimposed over two-color transmission plots for individual pixels
in $B$ and $I$,
in the zero-correction case. These point-by-point data are fitted to
derive the slope $R$ of the extinction curve (section 4).

We do not see a strong trend in arm/interarm extinction contrast with 
galactocentric radius manifested in these distributions. 
The uncorrected histograms in Fig. 9 show no connection between
extinction and position within the disk, while the maximally corrected ones
are more opaque within 2 kpc. However, these lanes are the ones
in which the foreground-light correction is large and uncertain, so it
is not clear whether the arm/interarm contrast changes with radius.
Spiral arms can include dust clouds 
on 15-pc scales with optical depths up to $\tau_B=1$ at any radius
within the disk. There are several dust patches far from the nucleus of
NGC 3314A whose allegiance is not obvious from the broad-band images,
but for which association with H$\alpha$ emission at its redshift
makes it clear that they are part of the foreground disk 
(examples are circled features 1, 7, and 8 in Fig. 6).

\subsection{H$\alpha$ Emission in the Inner Disk}

The fiber-array spectra show a remarkably number of detections
of H$\alpha$ and [N II] emission from the background system all
throughout the region behind NGC3314A. Even at the modest resolution
of our synthesized H$\alpha$ maps, there is a local anticorrelation between
foreground and background H$\alpha$ intensity, directly showing 
the connection between dusty regions and star formation. Furthermore,
at a given radius, the foreground H$\alpha$ intensity correlates with
dust lanes in absorption, supporting the same point.

We can use the intensity of H$\alpha$ from the background system
to obtain crude values for the area-weighted transmission
of light in these areas. Ryder \& Dopita (1994) found that, using
azimuthally averaged surface-brightness measures, typical spirals
have nearly the same surface-brightness slopes for H$\alpha$ and the $I$ 
continuum. Most of the galaxy-to-galaxy scatter is in the normalization of
the two, rather than slope differences, so we can apply this by 
normalizing the ratio in the outer, unobscured parts of NGC 3314B.
The different inclinations of the two galaxies mean that irregularities
in the ratio (fluctuations in H$\alpha$ equivalent
width with radius) will have their effect damped by contributions
from other regions at a range of radius in the background system but
similar radii in the foreground disk. Since the relation is
derived for the mean H$\alpha$ surface brightness in a disk annulus,
bright structures will not in themselves introduce a bias. The mean 
slope of their relation is given by 
$\mu_{H \alpha} = 0.64 \mu_I + {\rm constant}$
for surface brightnesses $\mu$.

The $I$ disk light from NGC 3314B was modeled as an exponential disk,
with slope taken from the regions which are seen free of significant
obscuration ($R > 24^{\prime\prime}$). Depending on how the elliptical 
measurement 
annuli are weighted,
a range of scale lengths $8.6-10.1^{\prime\prime}$ is found, which gives an
intensity 
difference of a factor 1.6 if both laws are normalized from
$24-30^{\prime\prime}$ and extrapolated inward to the nucleus. The H$\alpha$ 
deficit in each radial bin with respect to this model
was estimated by generating synthetic disks matching the expected
profiles based on the $I$ disk models, and dividing the observed
H$\alpha$ points by these models after convolving with a $3^{\prime\prime}$ 
flat-topped
filter to account for the fiber aperture size. Then the resulting
H$\alpha$ point-by-point transmission estimates were transformed into
a coordinate system in the plane of NGC 3314A and radially averaged.
This averaging ignored points without detected H$\alpha$, a point which
somewhat biases the results to areas of low extinction, though the
H$\alpha$ detections span the entire radial extent of NGC 3314A.
These data may be thought of as mostly an interarm measurement, since
the transmission measures are dominated by the clearest areas, although 
the filling factor of these areas enters as well. The derived
transmission values at the wavelength of H$\alpha$ are listed in Table 3,
ranging from 0.7 in the outer disk to 0.2 in the inner few arcseconds.
Here, the errors are dominated by uncertainties in the background
model, specifically the allowed range of $I$ surface brightness in
the inner disk predicted by the allowed fits in the outer portions.
This uncertainty dominates that from scatter in each annular 
averaging region and the scatter in the typical $I$-H$\alpha$ relation
from Ryder \& Dopita.

\subsection{Central Extinction}

The background nucleus of NGC 3314B is detected at $K$ and $I$ (the 
latter is particularly obvious in a color-composite display or $B-I$ color map).
We estimate the extinction in this direction 
(400 pc from the nucleus of NGC 3314A in its disk plane) by 
measuring the excess
flux from the nucleus over its surroundings (removing disk light
from NGC 3314A) and adopting typical $I-K$ colors for Sb nuclei.
In the $I$ image (expanded about the nucleus in Fig. 10), transmitted 
light from the background nucleus can be seen between several
discrete, filamentary dust lanes. Tracing along these relatively
transparent regions gives a surface brightness for transmitted light
there. The extinction we derive is large enough to be significant
even at $K$, so we make the mild assumption that the long-wavelength
reddening curve goes as $1/ \lambda$ so that $A_I = 2.7 A_K$ and adopt
a simple two-component model for the extinction. We assume that the
``transparent'' regions have constant extinction and that the arm regions
are completely opaque, using the $I$ image to give the fractional areas
of each kind within the seeing disk of the $K$ measurements at the
nucleus of NGC 3314B. Then the ratio of intensities at $I$ and $K$ can give
the extinction for the ``transparent'' regions. The ``opaque'' regions
allow less than one third of the transmission of the ``transparent''
ones, since the background nuclear light is not clearly detected 
through them.

An extremely red, near point-source object appears through one of the
deep dust lanes, at about the right position to be a central
intensity spike at the exact nucleus of
NGC 3314B. This contributes less than 10\% of the $I$ light from the
background nucleus (which we operationally take as light within a
radius of $r=1.1^{\prime\prime}$), and is unlikely to have such extreme colors 
as to seriously affect the $K$ flux and bias the color measurement.

The $I-K$ colors of galaxy nuclei should be robust, since the color evolution of
stellar populations is muted at these long wavelengths. Using an
old (15 Gyr) burst model from the Bruzual \& Charlot calculations
and the effective wavelengths of the F814W and $K$ filters gives
an unreddened color index $I-K = 1.64$. Within a radius $r=1.1^{\prime\prime}$
of the NGC 3314B nucleus, chosen to encompass the seeing disk of
the $K$ image, we derive a color for transmitted background light
of $I-K=4.05$ for a reddening $E_{(I-K)}=2.41$
and extinctions $A_I=3.30, A_K=1.22$. The larger extinctions in the
dust lanes crossing this region mean that light transmitted through them
can be ignored, even at $K$, at our levels of accuracy. Our color
could be in error at the 0.3 magnitude level due to the complex
structure and very different resolutions of the two images.
For reference, a typical Galactic extinction curve with this $I$ extinction 
has $A_B=6.9$, depending in any particular case on the filter shape
and background spectrum, since the reddening function is so steep.
In this case, extinction cannot be ignored near the nucleus even at 2.2$\mu$.

\subsection{A Composite Extinction Profile}

We summarize the results in this section by transforming the various
extinction measurements to inferred transmission at $B$, taking a
representative slope for the extinction curve as given by $R=3.1$
to transform data at other wavelengths. Fig. 11 shows all these
data in plots of $B$ transmission (linear and logarithmic)
versus radius within the disk
of NGC 3314A, including dust-lane and interlane measurements. This gives
an extinction picture of the spiral free of model assumptions as to
its own internal structure. More arm regions
clearly exist inside of 1.8 kpc with extinctions at least as
large, which we could not measure because of the lack of obvious interarm 
regions from which to derive background properties. 
The overall agreement between disparate techniques gives some confidence
in the general form of this plot, though the scatter near the nucleus
appears large in the logarithm (since the transmission is small, but
just how small depends on the spatial resolution and exact region sampled).
The smaller H$\alpha$ extinction near the nucleus may also
reflect the 3$^{\prime \prime}$ aperture size of the H$\alpha$ observations,
which may smear out the structure within this distance from the center.

\section{The extinction curve in NGC 3314A}

As in Keel \& White (2001), we examine the extinction curve by considering 
the relation between transmissions measured at $B$ and $I$ for sets
of points, with each slope (parameterized as is customary via
$R = A_V / E_{(B-V)}$) giving a different curve in a family. The
expected relation between transmissions in these passbands, $T_B$ and $T_I$,
was calculated by taking a model photon-rate spectrum for the background disk
(an old population with constant star-formation rate), folding this through 
the response functions of the filters and WFPC2 CCDs, applying a
reddening of known amount with each value of $R$, and deriving the
transmission values for each set of $A_V$ and $R$.

For each of the dust lanes in Table 2,
we derive the best-fitting
value of $R$ in a $\chi^2$ sense, with the typical error
derived from scatter in the upper region of each dust lane's data. 
Reference curves for the Galactic
mean $R=3.1$ are shown for each dust feature in Fig. 9. 
We also considered two ``control'' regions outside the obvious extent
of NGC 3314A, to show the effect of background structures on
derived extinctions after our interpolation procedure. The amount of
interlane extinction, with respect to which the differential dust lane values 
are measured, will not affect the measured slope of the
reddening law as long as there is not a systematic change in dust properties
between arm and interarm regions, and any such change will have its effects
muted since our estimates for interlane extinctions are much less than 
we see in the arm dust lanes.

Extinction curves with a variety of $R$ values ($R=1.1-5.1$) were fit
to the $I$ and $B$ transmission data for each region.
Errors for the individual pixel values
were estimated by determining the standard deviation
about the mean in a slice through the $I$ data
for a narrow range of $B$ transmission around $B=0.8$.
These errors were used in subsequent $\chi^2$ fits to
the distributions of $I$ versus $B$, both with no correction
and with maximal correction for foreground galaxy light (as in section 
3.2). Table 4 lists the best fit $R$ value for each region,
as well as $\chi^2_\nu$, the $\chi^2$ per degree of freedom.
In almost all cases, the errors are large enough, and the
$\chi^2$ distribution among fitted $R$ values so shallow,
that the full range of fitted $R$ values are consistent
(at the 90\% confidence level) with the minimum. However,
it is interesting that the scatter in fitted values is less than
this would suggest; in the uncorrected instance, they satisfy $R=3.7 \pm 0.8$
and in the maximum-correction case, $R=3.2 \pm 1.3$ (where the error
in each case is the standard deviation of the individual regions,
larger by a factor 2.2-3 than the standard deviation of the mean). Three 
regions
could not be fit in the maximum-correction data, as the scatter in
$B$ transmission from low-S/N pixels produced enormous scatter
in one axis for large $I$ transmissions. Since the form of the reddening
trajectory is tightly constrained and cannot follow most kinds of
scatter, the true errors may be smaller than quoted. Nevertheless,
the most we can state is that the reddening law in NGC 3314 is
broadly consistent with the Milky Way slope, and that there is no
strong trend in $R$ with radius in NGC 3314A.

This kind of analysis is limited to regions more than about 2 kpc from the
foreground nucleus. The combination of increasing extinction and brighter
foreground structure means that background features rapidly cease to
be identifiable inside this radius, although the H$\alpha$ results in 
section 3.3 show that significant light is still transmitted in
some areas.

\section{Dust Scale Height in the Disk of NGC 3314A}

Several large dust complexes in NGC 3314A are seen partially silhouetted by
the disk of NGC 3314B, providing an opportunity to test for the
signature of a thick dust lane. Numerous dust features in edge-on
spirals extend vertically beyond most of the disk stars
(e.g., Howk \& Savage 1999,2000), and the relative $z$-heights of dust and stars 
are important
in radiative-transfer models for galaxies. Our basic test is to
compare the colors of the dust lane when seen against the
background galaxy with those when it is seen by transmitted light from
within NGC 3314A itself. A physically thick dust layer will have
nearly the same color whether backlit by its own galaxy or another as
well, while a thin one will appear substantially less red when
not backlit externally. The use of color behavior to distinguish the
height of dust clouds from the plane (and implicitly their thickness,
as a population at any rate) was considered by Elmegreen (1980), who
presents calculations including $B-I$ change or various mixes of
cloud height and extinction. To first order, cloud height and thickness
are degenerate in the observed color, especially for large opacity.
In a simple model of well-mixed starlight and a geometrically thin
dust layer, the relative light deficit in a dust lane seen from
the foreground galaxy by itself will be half that of the same dust
seen against the background system as well, while the reddening
will be twice as large for a given light deficit. Since we deal with
relative amounts of ``missing" light, a strictly local comparison
independent of the amount of background intensity will suffice as long
as we consider regions where the background intensity is much higher than
the level of foreground light.

We tested for dust height using the large set of dust clouds seen
on and off the background disk including dust lane 4 from Fig. 6
and the features seen to its north against only the light from
the foreground spiral arm. For spatially confined dust, we
used a median-windowed version of the $I$ image (specifically the
result of dividing the image by a median-filtered image with a 
5.0$^{\prime \prime}$ window) to derive the relative intensity
of light received at each point, and compared these values to the $B-I$ colors
pixel by pixel. Histograms of the relative $I$ intensity showed
the backlit dust to have many pixels with values as low as 0.3, while
the non-backlit dust lanes have no values below 0.75, as predicted
for a thin dust lane. Comparable covering fractions are reached at,
for example, 0.4 for the backlit case and 0.65 for the non-backlit
areas. Similarly, the relative intensity-color
slope is much steeper for the non-backlit lane, such that the
changes in $B-I$ at transmission 0.75 are 0.08 and 0.20 magnitude
for the backlit and non-backlit regions, respectively. At least in
the regions amenable to this test, most of the dust must be
well-concentrated to the disk midplane.
 
\section{Conclusions}
 
We have used several techniques to measure the extinction in NGC 3314A,
using the backlighting provided by NGC 3314B. Such results should
be more robust than those from modelling a galaxy's extinction of its
own radiation, and certainly incorporate very different assumptions and 
sources of error.

Combining results from differential
extinction measures in dust lanes, color excess in interlane regions,
H$\alpha$ surface brightness of transmitted disk light, and the $I-K$
color of light from the background nucleus, we see a consistent
picture of dust extinction in this spiral. Converted to $A_B$ for
convenience, and not making a correction for disk inclination,
the extinction is essentially zero outside 0.7 $R_{25}$,
except for isolated and well-defined dust clumps whose densest
resolved regions reach $A_B=0.4$. Within this radius, the interarm
extinction rises by about $\Delta A_B = 0.12$ per kpc, and interspersed
dusty arms can include regions of high opacity $A_B > 1$. In the
innermost few hundred pc, even the most transparent regions between 
dust lanes show $A_B \approx 7$
and there are dusty arms with $A_B > 8.2$. These values correspond to
lines of sight all the way through the disk, and are thus at least
twice the effective extinction toward any particular component of
the galaxy as expressed in its own escaping light. For calculations
such as probabilities of absorption of QSO light, a good approximation
is that the transmission rises linearly with radius from the center, reaching
unity at about 0.7$R_{25}$ in blue light.

The dust in distinct arms shows a reddening behavior broadly consistent with
Galactic dust, perhaps to be expected for reasonably luminous spirals.
Fits by $\chi^2$ to two-color reddening trajectories give values for
the $R$ parameter 2.9--5.1, with a grand average from both data corrected
for maximal foreground light and uncorrected of $R = 3.5 \pm 0.3$
using the standard deviation of the mean.

One spiral arm shows comparable dust complexes in both backlit and
non-backlit regions. The color-intensity behavior in these regions
suggests that the dust is more strongly concentrated to the disk plane
than is the starlight, as found in the Milky Way and several external 
edge-on spirals.

\acknowledgements

We are grateful for the efforts and cooperation of the Hubble Heritage
Team, especially Lisa Frattare, in generating a data set of considerable
scientific as well as aesthetic value, partucularly in matching the
pointing and orientation of the two data sets so precisely. Ron Buta 
kindly made software
available that greatly speeded our new determination of these
galaxies' orientation parameters. We thank Barbara Cunow for
providing some of her data on $B-I$ color profiles in advance of publication.
This work was supported by NASA HST grant GO-06438.01-95A. The referee
made several suggestions which improved the clarity of our presentation,
which we and the reader appreciate.

\clearpage


\figcaption
{WFPC2 color composite image of NGC 3314. The image is in pixel coordinates,
oriented so that the top is $5.1 ^\circ$ west of north. This representation
was constructed using the summed $B$ data for blue, the mean of $V$ and $R$ 
data for green, and $I$ data as red. The nucleus of
NGC 3314B is visible as a heavily reddened object immediately to 
the west of the foreground nucleus of NGC 3314A. The region shown
subtends a $635 \times 700$-pixel region ($63 \times 70^{\prime\prime}$)
from WFPC2 CCD chip WF3.
\label{fig1}}

\figcaption
{K-band IRTF image of NGC 3314. North is at the top, and the field
shown is 60 arcseconds (200 pixels) on a side, with north at the top. 
The background nucleus of NGC 3314B, and the distinct
spiral patterns of both galaxies, are prominent.
\label{fig2}}

\figcaption
{Velocity field of NGC 3314A from WIYN H$\alpha$ measurements. The individual
fiber detections have been gridded onto a $1^{\prime\prime}$ mesh by averaging 
all detections
covering a given synthetic pixel. The gray scale indicates pixel-by-pixel values
following the scale at the bottom, and contours are superimposed at
intervals of 50 km s$^{-1}$. The nucleus lies at the northwest edge of
the data gap near the center of the disk.
\label{fig3}}

\figcaption
{Velocity field of NGC 3314B from WIYN H$\alpha$ measurements, constructed
as in Fig. 3. The nucleus is at the western side of the central
``hole" in the map. As the scale bar shows, emission is detected 
over a much wider area from NGC 3314B than from NGC 3314A, with
some of the foreground dust lanes visible as gaps in the
background velocity field.
\label{fig4}}

\figcaption
{Fits to the individual rotation curves of NGC3314A and 3314B, generated from
the velocity fields in Figs. 3--4 under the assumption of pure circular
motion. Error bars show the standard deviation of each mean based on individual
measurements, increasing to the center (especially for NGC 3314B) 
because of the small number of points, so that the precise center of
the velocity field may be poorly determined or biased by the 
distribution of H$\alpha$ detections. The radius in each case is in arcseconds,
as projected along the major axis at each component galaxy's distance.
\label{fig5}}

\figcaption
{Interlane regions measured for extinction via $B-I$ color
excess, as listed in Table 1. They are marked on the $B$-band WFPC2
image, shown with a pseudologarithmic intensity scale to compress
the dynamic range while avoiding the distracting amplification
of sky noise. Also marked are discrete dust lanes measured for 
for differential extinction against their surroundings
(Table 2), indicated by circled numbers. Features 1, 7, and 8 are unambiguously 
associated with the foreground system by showing H$\alpha$ emission at its
redshift, despite their large projected distance.
\label {fig6}}

\figcaption
{Measured $B-I$ color in ``transparent'' interarm regions versus
radius in both foreground and background disks. The relation in the background
disk can be used to limit an internal, intrinsic $B-I$ gradient, shown
as the dotted line passing through the lower points. The stronger 
slope against radius in the foreground disk is interpreted as
a reddening signature, since a simple broken linear fit (dotted line)
has a much stronger slope for $R<4$ kpc.
 \label{fig7}}

\figcaption
{Our procedure for deriving extinctions in discrete dust lanes with
respect to their immediate surroundings. This is illustrated with a
section of the $B$ image encompassing feature 1 from Fig. 6. An area
including the dust feature and nearby background spiral structure
(a) is rotated (in this case by $51^\circ$)
to align the $x$-axis with most of the background spiral
pattern (b). Then the spiral pattern is interpolated along this
pitch angle (c), avoiding bright associations. This interpolated image 
is rotated back to the observed frame (d), where steps (b)-(d) are
carried out on images expanded by a factor 2 to reduce numerical
effects of interpolation. Correction for foreground light 
amounts to a constant subtracted from (d).
Finally, the original image is divided by
the interpolated model to give a map of estimated transmission
(e). There is only minor streaking along the direction of interpolation
in (d), suggesting that the reconstruction is not strongly influenced by 
details of the masking and interpolation in (c).
The darkest areas in (e) have a $B$ transmission near 0.40.
Similar analysis was done for each dust lane, in both $B$ and $I$.
 \label{fig8}}

\figcaption
{Montage of B versus I transmission data for the dust lanes listed in
Table 2, plotted without correction for foreground light. The solid
and dashed curves are marginal distributions of blue transmission
for no foreground correction (solid) and maximum foreground correction
(dashed), which must in some cases be a substantial overcorrection
since this leads to large numbers of nonphysical negative values.
These are shown binned every 0.05 in transmission, with normalization
which is arbitrary but the same for both distributions for a given
dust lane.
\label {fig9}}
 
\figcaption
{Detail of the WFPC2 $I$ image expanded around the nuclei. The contours
show the brightest isophotes from the $K$ image in
Fig. 2, showing the secure detection of excess $I$ light 
closely coincident with the $K$ peak of NGC 3314B. The dashed ellipse
shows the region within which background nuclear light was analyzed. 
This $9.6 \times 10.2$-arcsecond region is oriented as in Fig. 1,
which shows the heavily reddened color of the background light 
in this area.
\label {fig10}}

\figcaption
{Composite extinction profile, plotted in both linear transmission (a)
and logarithmically, to stress the further depth of extinction near
the nucleus (b). This combines the $I-K$ data on the background
nucleus (large circle at $r=0.4$ kpc), the $B-I$ interarm data (open circles), 
the inner-disk mean extinction from H$]alpha$ surface brightness 
(filled circles), and point-by-point measures in dust lanes referenced to
the running interarm model (small points). The dust lane data are shown
with the proper radius and width, but no attempt was made to plot
individual points at their respective radii. Additional dust lanes exist
throughout the inner disk, but we can show only those which the
surrounding and backlighting allowed us to measure. 
\label {fig11}}

\clearpage

\begin{table*}
\begin{center}
\begin{tabular}{rrrrrc}
\tableline
\tableline
ID & $\Delta \alpha$ (s) & $\Delta \delta$ ($^{\prime \prime}$) & R(bg) (kpc) 
& R(fg) (kpc) & $B-I$\\   
\tableline
 1  & -1.27 &  18.2 & 5.51  &  4.18  &  $1.29 \pm 0.08$\\
 2  & -1.20 &  15.1 & 4.91  &  3.81  &  $1.18 \pm 0.07$\\
 3  & -0.84 &  18.4 & 5.16  &  3.38  &  $1.02 \pm 0.12$\\
 4  & -0.84 &  13.9 & 3.99  &  2.92  &  $1.38 \pm 0.14$\\
 5  & -0.68 &  12.1 & 3.42  &  2.43  &  $1.72 \pm 0.10$\\
 6  & -0.37 &  13.9 & 4.16  &  2.27  &  $1.07 \pm 0.13$\\
 7  & -0.72 &   3.3 & 3.04  &  2.12  &  $1.85 \pm 0.15$\\
 8  & -0.67 &  -2.7 & 4.18  &  2.26  &  $1.27 \pm 0.15$\\
 9  &  0.42 &  -9.2 & 2.59  &  1.69  &  $1.31 \pm 0.24$\\
10  &  0.31 & -12.0 & 3.61  &  1.95  &  $1.22 \pm 0.24$\\
11  &  0.52 & -10.3 & 2.89  &  1.93  &  $1.39 \pm 0.27$\\
12  &  0.62 & -10.4 & 2.98  &  2.16  &  $1.64 \pm 0.33$\\
13  &  0.75 &  -7.1 & 2.89  &  2.25  &  $1.24 \pm 0.13$\\
14  &  0.75 & -11.4 & 3.37  &  2.51  &  $1.16 \pm 0.07$\\
15  &  0.47 & -13.0 & 3.72  &  2.21  &  $1.24 \pm 0.10$\\
16  &  0.41 & -16.1 & 4.85  &  2.61  &  $1.35 \pm 0.08$\\
17  &  0.96 & -12.8 & 4.02  &  3.08  &  $1.36 \pm 0.10$\\
18  &  0.72 & -16.2 & 4.53  &  2.92  &  $1.34 \pm 0.08$\\
19  &  0.70 & -19.2 & 5.48  &  3.29  &  $1.12 \pm 0.08$\\
\tableline
\end{tabular}

\tablenum{1}
\caption{
Selected interarm regions and measured colors \label{tbl1}}
\end{center}

\end{table*}

\begin{table*}
\begin{center}
\begin{tabular}{rrrr}
\tableline
\tableline
ID & $\Delta \alpha$ (s) & $\Delta \delta$ ($^{\prime \prime}$) &  R(kpc)\\
\tableline
1 & 0.89 & -20.9 & 3.74\\
2 & 0.56 & -13.8 & 2.43\\
3 & 0.34 & -15.8 & 2.55\\
4 & 0.59 &  -6.1 & 1.80\\
5 & 0.29 & -12.9 & 2.08\\
6 & 0.07 & -15.6 & 2.57\\
7 & -1.29 &   7.5 & 3.81\\
8 & -1.15 & 10.1 & 3.43\\
9 & -0.27 & 10.4 & 1.69\\
\tableline
\end{tabular}

\tablenum{2}
\caption{
Discrete Dust Lanes Fitted for Extinction Curves\label{tbl2}}
\end{center}

\end{table*}

\begin{table*}
\begin{center}
\begin{tabular}{cc}
\tableline
\tableline
Radius ($^{\prime \prime}$) & Transmission \\
\tableline
1    &   $0.22 \pm 0.07$\\  
3    &   $0.38 \pm 0.11$\\   
6    &   $0.42 \pm 0.11$\\   
9    &   $0.56 \pm 0.12$\\ 
12   &   $0.56 \pm 0.10$\\ 
15   &   $0.63 \pm 0.12$\\	 
18   &   $0.76 \pm 0.12$\\  
21   &   $0.88 \pm 0.11$\\  
24   &   $0.92 \pm 0.12$\\  
\tableline
\end{tabular}

\tablenum{3}
\caption{
H$\alpha$ extinction measures in $1-24^{\prime\prime}$ annuli \label{tbl3}}
\end{center}

\end{table*}
\clearpage

\begin{table*}
\begin{center}
\begin{tabular}{ccccc}
\tableline
\tableline
  & \multicolumn{2}{c}{ uncorrected } & \multicolumn{2}{c}{ corrected } \\
Region & $R$ & $\chi^2_\nu$  & $R$ & $\chi^2_\nu$  \\
\tableline
1  &   2.9 &  0.62  &   ... &  ...   \\
2  &   3.5 &  0.93  &   ... &  ...   \\
3  &   3.3 &  0.65  &   ... &  ...   \\
4  &   3.9 &  0.96  &   1.9 &  1.57  \\
5  &   5.1 &  0.77  &   4.5 &  0.98  \\
6  &   3.3 &  1.11  &   2.7 &  1.30  \\
7  &   3.5 &  1.00  &   2.9 &  1.16  \\
8  &   3.3 &  0.92  &   2.1 &  1.08  \\
9  &   5.1 &  0.92  &   5.1 &  1.11  \\

\tableline
\end{tabular}

\tablenum{4}
\caption{
Best-fitting $R$ values for dust lanes \label{tbl4}}
\end{center}

\end{table*}

\end{document}